\documentclass[12pt]{article}
\usepackage{epsfig}
\def\wtG{\widetilde G}
\def\ovl{\overline }
\def\rlh{\rightleftharpoons}

\begin{document}

\begin{flushright}
hep-ph/0209153\\
\end{flushright}

\begin{center}
{\Large\bf
Mixing and decays of $\rho$- and $\omega$-mesons
}

\vspace{0.4cm}

{\large Ya.I.Azimov \footnote{e-mail: azimov@pa1400.spb.edu}}

\vspace{0.3cm}

{\it Petersburg Nuclear Physics Institute,\\
Gatchina, St.Petersburg, 188300, Russia}

\vspace{0.3cm}
\end{center}

\begin{abstract}
Isospin violating mixing of $\rho$- and $\omega$-mesons
is reconsidered in terms of propagators. Its influence
on various pairs of $(\rho^0,\,\omega)$-decays to the same
final states is demonstrated.
Some of them, $(\rho^0,\,\omega)\to\pi^+\pi^-$ and
$(\rho^0,\,\omega)\to\pi^0\gamma$, have been earlier discussed
in the literature, others ({\it e.g.}, $(\rho^0,\,\omega)\to
\eta\gamma$ and $(\rho^0,\,\omega)\to e^+e^-$) are new in this
context. Changes in partial widths for all the decay pairs are
shown to be correlated. The set of present experimental data,
though yet inconclusive, provides some limits for the direct
$(\rho\omega)$-coupling and indirectly supports enhancement of
$\rho^0\to\pi^0\gamma$ in comparison with
$\rho^{\pm}\to\pi^{\pm}\gamma$, though  not so large as in some
previous estimates.

\vspace{0.3cm}
PACS numbers: 11.30.Ly, 13.25.Jx, 14.40.Cs

\end{abstract}

\section {Introduction}

Isospin conservation was considered for many years to be a good symmetry
of strong interactions, though violated due to electromagnetic (e.m.)
corrections. Of course, e.m. mechanism of isospin violation should
exist. However, the quark picture and QCD have suggested one more
interesting possibility, to violate isospin by strong interactions
as well. This is possible due to mass difference between $u$ and $d$
quarks. Parametricaly, such mechanism could be stronger than e.m. one,
but its exact value essentially depends on unknown hadronic matrix
elements and might appear numerically suppressed, at least, in some
cases. In reality, for most of known manifestations the violation may
look quantitatively compatible with pure e.m. nature (numerically they
are $\sim{\cal O}(\alpha)$ in amplitudes, or even smaller).  Therefore,
very elaborate work, both theoretical and experimental, will be
necessary to pick out the underlying physics and separate various
sources of isospin violation.

A special situation appears in the decay $\omega\to\pi^+\pi^-$
(branching ratio about 2\% \cite{PDT2}), where enhancement becomes
possible (and seems to be operative) due to transition of $\omega$
into $\rho^0$ having the near mass and large width
$\rho^0\to\pi^+\pi^-$. However, experiments with this mode can
extract only one real parameter (instead of two or more, see below
for details) and, thus, are insufficient to reveal the isospin
violation mechanism(s).

In this respect, one more decay mode has attracted much attention in
recent years. It is the radiative decay $\rho^0\to\pi^0\gamma$. Its
partial width was expected to be the same as for the charged companion
$\rho^\pm\to\pi^\pm\gamma$.  Meanwhile, experiment seems to give
evidence \cite{PDT2} for a higher branching ratio of the neutral mode
as compared to the charged one, though the result might still change
\footnote{One should make some reservations here.  Measurement methods
are very different:  Primakoff effect for $\rho^\pm$, and
$e^+e^-$-annihilation for $\rho^0$.  Backgrounds are also very
different and not quite clear for $\rho^0$, see short discussion
below.}.  Qualitatively, it may have reasonable explanation as being
due to mixture of the direct decay and the cascade transition through
$\omega$ with the relatively large amplitude $\omega\to\pi^0\gamma$.

Quantitative consideration has been mainly done in terms of a kind
of effective field theory with some model Lagrangian (like Vector
Dominance Model, Chiral Perturbation Theory and so on, see
ref.\cite{BOC} and many references therein). Such approaches, to
be applicable, need some limitations ({\it e.g.}, constant and real
transition vertices), which may appear too restrictive. Another
approach, in terms of propagators, was applied more recently~\cite
{WJY}.  Motivated by summing general Feynman graphs, it gave
unexpectedly large enhancement for $\rho^0\to\pi^0\gamma$.

In the present paper this approach is reconsidered more accurately.
The consideration is then extended further to show that the
$(\rho\omega)$-mixing should affect a wider set of decay modes
where effects of mixing should be possible as well providing
enhancements or suppressions of partial widths. Indeed, the present
data qualitatively confirm the expected role of mixing. Such a way,
at better experimental accuracy, may help to construct a consistent
picture of the isospin violation and to clarify its nature. What
about the enhancement suggested in ref.\cite{WJY}, it is shown
to be strongly overestimated.

\section {Propagator description for mixing of vector particles}

The unperturbed propagator for a vector meson $V$ with "bare" mass
$M_V^{(0)}$ may be presented in the form
\begin{equation}
[D_V^{(0)}(k^2)]_{\mu \nu}=
\frac{g_{\mu \nu}-\frac{k_\mu k_\nu}{M^{(0)2}_V}}
{k^2-M^{(0)2}_V} =
\frac1{k^2-M^{(0)2}_V}\,
\left(g_{\mu \nu}-\frac{k_\mu k_\nu}{k^2}\right)-
\frac1{M^{(0)2}_V}\,
\frac{k_\mu k_\nu}{k^2}
\,.
\end{equation}
Note that $(g_{\mu \nu}-\frac{k_\mu k_\nu}{k^2})$ and
$\frac{k_\mu k_\nu}{k^2}$ provide a complete set of
projectors since
$$\frac{k_\mu k_\alpha}{k^2}\,g^{\alpha \beta}\,
\frac{k_\beta k_\nu}{k^2}=
\frac{k_\mu k_\nu}{k^2}\,,~~~~
\left(g_{\mu \alpha}-\frac{k_\mu k_\alpha}{k^2}\right)\,
g^{\alpha \beta}\,
\left(g_{\beta \nu}-\frac{k_\beta k_\nu}{k^2}\right)=
\left(g_{\mu \nu}-\frac{k_\mu k_\nu}{k^2}\right)
\,, $$
\begin{equation}
\left(g_{\mu \alpha}-\frac{k_\mu k_\alpha}{k^2}\right)\,
g^{\alpha \beta}\,
\frac{k_\beta k_\nu}{k^2}=0
\,,~~~~
\left(g_{\mu \nu}-\frac{k_\mu k_\nu}{k^2}\right)+
\frac{k_\mu k_\nu}{k^2}=g_{\mu \nu} \,.
\end{equation}
The most general form of the vertex for two vector meson
transition $V_1\to V_2$ also contains two terms, transversal
and longitudinal:
\begin{equation}
[\Gamma_{12}(k^2)]_{\mu\nu}=G_{12}(k^2)\,
\left(g_{\mu \nu}-\frac{k_\mu k_\nu}{k^2}\right)+
F_{12}(k^2)\,\frac{k_\mu k_\nu}{k^2}\,.
\end{equation}
The vertex for the transition $V_2\to V_1$ is similar. Moreover,
$T$-invariance makes it just the same. We retain, however, formal
difference of, say, $G_{12}$ and $G_{21}$ to reveal the structure
of arising expressions.

Now we can describe evolution of any initial state. The full
propagator for $V_1\to V_1$ is
$$
D_{11}= D_1^{(0)}+D_1^{(0)}\Gamma_{12}D_2^{(0)}\Gamma_{21}D_1^{(0)}+
D_1^{(0)}\Gamma_{12}D_2^{(0)}\Gamma_{21}D_1^{(0)}\Gamma_{12}D_2^{(0)}
\Gamma_{21}D_1^{(0)}+...\,. $$
The summation runs separately for each of the projector terms due to
their orthogonality, so we obtain
\begin{equation}
[D_{11}]_{\mu\nu}=(k^2-M_2^{(0)2})\,R_t(k^2)\,
\left(g_{\mu \nu}-\frac{k_\mu k_\nu}{k^2}\right)-
M_2^{(0) 2}\,R_l(k^2)
\, \frac{k_\mu k_\nu}{k^2}
\end{equation}
with
$$
R_t(k^2)=[(k^2-M_1^{(0)2})(k^2-M_2^{(0)2})-
G_{12}G_{21}]^{-1}\,,
$$
\begin{equation}
R_l(k^2)=[M_1^{(0)2}M_2^{(0)\,2}-F_{12}F_{21}]^{-1}\,.
\end{equation}
The full propagator for the transition $V_1\to V_2$ takes the form
\begin{equation}
[D_{12}]_{\mu\nu}=G_{12}\,R_t(k^2)\,
\left(g_{\mu \nu}-\frac{k_\mu k_\nu}{k^2}\right)+
F_{12}\,R_l(k^2)
\, \frac{k_\mu k_\nu}{k^2}\,.
\end{equation}
Full propagators for transitions $V_2\to V_2$ and $V_2\to V_1$ can be
obtained from eqs.(4),(6) by interchange of the indices 1 and 2. In
all the expressions one may, generally, consider $M_1^{(0)},\,
M_2^{(0)}$ to be also $k^2$-dependent. The above description can be
applied to mixing of any vector mesons ({\it e.g.}, $\varphi$ and $\omega$).
It may even be generalized to mixing of any number of mesons (say,
$\rho$-$\omega$-$\varphi$, or admixture of radially excited states).

For the particular case of $(\rho,\omega)$-mixing the formulae
simplify. It is, first of all, due to nearness of the "bare" masses:
$\left|\delta M^2/M^2\right|\sim 0.1\,$, where
\begin{equation}
\delta M^2=\frac{M^{(0)2}_\omega-M^{(0) 2}_\rho}2\,,
~~~M^2=\frac{M^{(0)2}_\omega+M^{(0)2}_\rho}2\,,
\end{equation}
\begin{equation}
M^{(0)}_{\omega,\,\rho}=m^{(0)}_{\omega,\,\rho}-
\frac{i}2\,\Gamma^{(0)}_{\omega,\,\rho}
\end{equation}
(we take masses and widths from PDTables \cite{PDT2}).
As a result, the essential $k^2$-region is small, in
vicinity of the masses, and that allows one to consider the transition
vertices as constants: say, $G(k^2)\to G(M^2)$. The constancy of
vertices corresponds to what is assumed in effective field theories.
However, the effective vertex $G$ may appear complex, while it
should be real for self-consistency of field theory. (More exactly,
in the field theory one should be able to change the phase of $G$
by rephasing the fields $\omega$ and $\rho\,$; in this way one can
make arg$\,G=0\,$. If, however, $G$ contains contributions of real
intermediate states, such as  $2\pi\,$ and $3\pi$, then the
\mbox{$\omega\rho$-rephasing} may be insufficient to assure real
$G\,$.) Corrections for $k^2$-dependence, when taken in the framework
of an effective field theory, may also provide difficulties.

At constant vertices, the longitudinal part $R_l$ has no poles in
$k^2$ (and no $k^2$-dependence at all), while $R_t$ has two poles
corresponding to "physical" states $\rho$ and $\omega$ ({\it cf}.
with the structure of unmixed propagator (1) ).  The "physical"
masses are equal
\begin{equation} M_\omega^2= M^2+K\delta M^2\,,~~~~ M_\rho^2=
M^2-K\delta M^2\,, \end{equation}
where
\begin{equation}
K=\sqrt{1+\wtG_{\rho\omega}\wtG_{\omega\rho}}\,,~~~
\wtG_{\rho\omega}=\frac{G_{\rho\omega}}{\delta M^2}\,,~~
\wtG_{\omega\rho}=\frac{G_{\omega\rho}}{\delta M^2}\,.
\end{equation}

Let us consider a process $i\to f$ with intermediate
$\rho$- and $\omega$-mesons. Its amplitude is
\begin{equation}
A_{if}=A^{(0)}_{i\rho} D_{\rho\rho}A^{(0)}_{\rho f}\,+\,
A^{(0)}_{i\rho} D_{\rho\omega}A^{(0)}_{\omega f}\,+\,
A^{(0)}_{i\omega} D_{\omega\omega}A^{(0)}_{\omega f}\,+\,
A^{(0)}_{i\omega} D_{\omega\rho}A^{(0)}_{\rho f}\,,
\end{equation}
where $A^{(0)}_{i\rho}, A^{(0)}_{i\omega}$ are production
amplitudes for "bare" $\rho$-, $\omega$-states, while
$A^{(0)}_{\rho f}, A^{(0)}_{\omega f}$ are their decay amplitudes.
We will be really interested here in decay modes ($e^+e^-,\,
\pi^0\gamma$ and some others) with current conservation, thus
only transversal parts of propagators are operative. Then we can
rewrite the amplitude through contributions of the "physical" states
\begin{equation}
A_{if}=A_{i\rho} D_{\rho}A_{\rho f}\,+\,
A_{i\omega} D_{\omega}A_{\omega f}\,,
\end{equation}
with "physical" propagators
\begin{equation}
[D_\rho(k^2)]_{\mu \nu}=
\frac{g_{\mu \nu}-\frac{k_\mu k_\nu}{M^{2}_\rho}}
{k^2-M^{2}_\rho} \,,~~~
[D_\omega(k^2)]_{\mu \nu}=
\frac{g_{\mu \nu}-\frac{k_\mu k_\nu}{M^{2}_\omega}}
{k^2-M^{2}_\omega}
\end{equation}
(see eq.(9) for masses) and "physical"
amplitudes
\begin{equation}
A_{i\rho}=\sqrt{\frac{K+1}{2K}}\left(A^{(0)}_{i\rho}-
A^{(0)}_{i\omega}\frac{\wtG_{\omega\rho}}
{K+1}\right)\,,~~~
A_{i\omega}=\sqrt{\frac{K+1}{2K}}\left(A^{(0)}_{i\omega}+
A^{(0)}_{i\rho}\frac{\wtG_{\rho\omega}}
{K+1}\right)
\end{equation}
for the meson production and
\begin{equation}
A_{\rho f}=\sqrt{\frac{K+1}{2K}}\left(A^{(0)}_{\rho f}-
\frac{\wtG_{\rho\omega}}{K+1}
A^{(0)}_{\omega f}\right)\,,~~~
A_{\omega f}=\sqrt{\frac{K+1}{2K}}\left(A^{(0)}_{\omega f}+
\frac{\wtG_{\omega\rho}}{K+1}
A^{(0)}_{\rho f}\right)
\end{equation}
for meson decays. The structure of masses (9) and amplitudes
(14),(15) corresponds to diagonalizing  the mass squared
matrix of the $(\rho,\omega)$-system
\begin{equation}
{\cal M}^2=
\left(\begin{array}{cc}
M_\rho^{(0)\,2}& G_{\omega\rho}\\
G_{\rho\omega}&M_\omega^{(0)\,2}
\end{array}\right)
\end{equation}
and its matrix propagator ${\cal D}= (k^2-{\cal M}^2)^{-1}$
in the form
\begin{equation}
{\cal M}^2=
\sqrt{ \frac{K+1}{2K}}
\left(\begin{array}{cc}
1&\frac{\wtG_{\omega\rho}}{K+1}
\\
-\frac{\wtG_{\rho\omega}}{K+1}
&1\end{array}\right)  \cdot
\left(\begin{array}{cc}
M_\rho^{2}&0\\0&
M_\omega^{2}\end{array}\right) \cdot
\sqrt{ \frac{K+1}{2K}}
\left(\begin{array}{cc}
1&
-\frac{\wtG_{\omega\rho}}{K+1}
\\
\frac{\wtG_{\rho\omega}}{K+1}
&1\end{array}\right)\,.
\end{equation}

Additional simplifications arise from $T$-invariance which allows
one to choose phases of states so that
$G_{\rho\omega}=G_{\omega\rho}\,$. Further, from previous
experience of isospin violation we expect the
$(\rho\omega)$-transition vertices to be numerically small.
{\it E.g.}, the e.m. mechanism gives
$\,|G|,|F|\sim {\cal O}(\alpha)\cdot M^2\sim 10^{-2}\,M^2\,$.
Then $|\wtG_{\rho\omega}|\sim 0.1\,$. In such a case
$|K-1|\sim10^{-2}\,,$ and with sufficient accuracy we can
substitute $K=1$ in eqs.(14),(15),(17).  Corrections for deviation
of $K$ from unity correspond to accounting for cascade returns
$\,\rho\to\omega\to\rho\,$ and/or $\,\omega\to\rho\to\omega\,$.

The above picture of $(\rho\omega)$-mixing is quite similar
to the well-known mixing of $(K^0\ovl K{}^0)$ as described
by Lee, Oehme, Yang \cite{LOY}. The bare states $|\rho^{(0)}>$
and $|\omega^{(0)}>$ appear to be analogs of $|K{}^0>$ and
$|\ovl K{}^0>$, while the physical states
\begin{equation}
|\rho>=N_\rho\left(|\rho^{(0)}>-\,
\frac{\wtG_{\rho\omega}}{K+1}
\,|\omega^{(0)}>\right)\,,~~~
|\omega>=N_\omega
\left(\frac{\wtG_{\omega\rho}}{K+1}
\,|\rho^{(0)}>+\,|\omega^{(0)}>\right)
\end{equation}
play the role of $|K_S>$ and $|K_L>\,$ (compare with expressions
(15); $N_\rho$ and $N_\omega$ are normalizing factors).
The essential difference, however, is $\delta M^2\neq0$, which
would imply $CPT$-violation in the case of $(K^0\ovl K{}^0)$.

This similarity reveals one more property of the
$(\rho\omega)$-system. In the case of exact isospin conservation
the bare states $\rho^{(0)}$ and $\omega^{(0)}$ can not be coherent,
and their phases (absolute and/or relative) are totally independent.
If mixing is possible, the physical states $\rho$ and $\omega$ can
be coherent to each other, so their relative phase becomes
physically meaningful. Nevertheless, phases of the bare states stay
arbitrary and may be changed so to not change phases of the physical
states (of course, phases of the normalizing factors and of
$\wtG_{\omega\rho},\,\wtG_{\rho\omega}$ should be changed
correspondingly). Such procedure, rephasing, is familiar in
description of neutral kaons, with only rephasing-invariant
quantities being physically meaningful and measurable. It may be
useful also for the $(\rho\omega)$-system.

For the $(K^0\ovl K{}^0)$-system with $CPT$-conservation we know
that the states $|K_S>$ and $|K_L>\,$ would be mutually orthogonal
only if $T$ (or $CP$) were conserved. The bare states
$\,|\rho^{(0)}>\,$ and $\,|\omega^{(0)}>\,$ in the
$(\rho\omega)$-system  are, surely, orthogonal to each other,
but the physical states $\,|\rho>\,$ and $\,|\omega>\,$ can be
also orthogonal only if
\begin{equation}
\frac{\wtG_{\rho\omega}}{K+1}= \frac{\wtG_{\omega\rho}^*}{K^*+1}\,.
\end{equation}
Evidently, this condition implies
$|\wtG_{\rho\omega}|=|\wtG_{\omega\rho}|\,$
({\it i.e.}, $\,|G_{\rho\omega}|=|G_{\omega\rho}|\,$), which
would be provided by $T$-invariance. Hence, the $T$-invariance
is necessary for $(\rho,\omega)$-orthogonality, in similarity
with neutral kaons. It is not, however, sufficient. Eq.(19) is
consistent with definition (10) only if $K$ (and
$\wtG_{\rho\omega}\wtG_{\omega\rho}$) is real. Note that
combinations $\wtG_{\rho\omega}\wtG_{\omega\rho}$ (and $K$
as well) and $\wtG_{\rho\omega}/\wtG_{\omega\rho}^*$ are
rephasing-invariant, {\it i.e.}, not changed if phases of
$\,|\rho^{(0)}>\,$ and $\,|\omega^{(0)}>\,$ change under fixed
phases of $\,|\rho>\,$ and $\,|\omega>\,$. Thus, the necessary
and sufficient condition of the $(\rho,\omega)$-orthogonality
is possibility to choose such phases of the bare states that
$\wtG_{\rho\omega}$ and $\wtG_{\omega\rho}$ are equal and real.
This condition may appear not natural (see discussion below and
recall that the $\wtG$'s contain the complex denominator $\delta M^2$),
so, most probably, the mixed physical eigenstates are non-orthogonal
even in spite of the $T$-conservation.

There is one more similarity between $(\rho\omega)$ and
$(K^0\ovl K{}^0)$ systems. $CP$-violation for neutral
kaons can manifest itself in two forms: the mixing violation
due to a $CP$-violating structure of the kaon effective
Hamiltonian, and the direct violation related directly to
kaon decay amplitudes. Isospin violation in the
$(\rho\omega)$-system may, analogously, have two forms: the
mixing violation due to isospin-violating structure of the mass
squared matrix (16) (nonvanishing $G_{\rho\omega}$  and/or
$G_{\omega\rho}\,$), and the direct violation (nonvanishing,
even if being isospin-forbidden, amplitudes $A^{(0)}_{\rho},
A^{(0)}_{\omega}$ for the bare meson production and/or decay
amplitudes).

If we could produce pure states $\rho^{(0)}$, $\omega^{(0)}$
and observe their decays in real time, we would see oscillating
time distributions (analogues of oscillating decays of pure $K^0$
and $\ovl K{}^0$). But this is surely unrealistic because of
too short lifetimes, and experiments can study only two-pole
structure of time-integrated $k^2$-dependencies. It should be
emphasized that in any experiment one can extract only those
poles related to physical $\rho$-,$\,\omega$-states, with
residues containing physical amplitudes (14),(15). The bare
states $\rho^{(0)},\,\omega^{(0)}$ and their amplitudes are
unobservable.

The latter discussions in terms of states has implicitly assumed
that bare amplitudes, bare masses and vertices
$G_{\rho\omega},\,G_{\omega\rho}$ are constants. However, such
assumptions are not necessary. All expressions (7)-(17) conserve
their form if (all or some of) the above quantities depend on
$k^2\,$. Then the mixing parameters $\wtG_{\rho\omega},\,
\wtG_{\omega\rho},\,K,\,$ as well as the "physical amplitudes"
and "physical masses" (of course not pole ones) become also
$k^2$-dependent.

\section {$(\rho\,\omega)$-mixing in particular decay modes }

The most popular in the literature on $(\rho\,\omega)$ isospin
violation is the meson mixing described by parameters $\wtG$
since they reveal some enhancement due to the small value of
$\delta M^2$ in denominator, see definition (10). With the
reasonable assumption of $T$-invariance we can choose phases
of the bare states so to have one (generally, complex) dimensionless
parameter $\wtG_{\rho\omega}=\wtG_{\omega\rho}\,$, which is universal
in all particular processes. In an effective field theory the related
parameters $G_{\rho\omega}, G_{\omega\rho}$ appear in the Lagrangian
as coupling constants for direct transitions $\omega\!\rlh\!\rho\,$.
Due to Hermiticity, they may be taken equal and real (the corresponding
terms by themselves are inevitably $T$-invariant). Even in this case
the complexity of $\wtG_{\rho\omega}=\wtG_{\omega\rho}\,$ can not be
removed; it is totally determined by complexity of $\delta M^2$, due
to widths of $\rho$ and $\omega$. Note that for the current
experimental values of masses and widths~\cite{PDT2}
$$2\delta M^2=M_\omega^2-M_\rho^2=(23368 + 108443i)\,{\rm MeV^2}\,,$$
{\it i.e.}, $\delta M^2$ is mainly imaginary, due to the large
$\Gamma_\rho$.

Realistic situation may be different. Transitions
$\omega\!\rlh\!\rho$ may go through some intermediate states,
virtual or real. If only virtual states are possible (say, for
the transition $\omega\to K\bar K\to\rho$ advocated
in \cite{BOC}), then $G_{\rho\omega}= G_{\omega\rho}$ are pure
real indeed. However, if real intermediate states (say, pions
or pions with one photon) are also essential, then
$G_{\rho\omega}= G_{\omega\rho}$ should be complex by themselves.
Correspondingly, parameters $\wtG_{\rho\omega}=\wtG_{\omega\rho}\,$,
being also universal, should have additional complexity, not related
to $\delta M^2$. Such a case is surely out of framework of an
effective field theory.

Apart from mixing, the isospin violation can manifest itself
directly in amplitudes of production and/or decay of bare states
$\,\rho^{(0)}\,$ and $\,\omega^{(0)}\,$ (see eqs.$(14),(15)\,$).
Intuitively, such contributions have no enhancement and should be
smaller than the enhanced mixing violation of isospin. However, this
may appear not universally true. In particular, effective mechanisms
for direct and mixing violations may appear different. This could
make the direct isospin violation be essential in some processes,
though negligible in others (again, phenomenologically similar to
apparent properties of $CP$-violation).

In any case, at the present state of knowledge and experience
one needs to use some model assumptions on the amplitudes and
mixing. That is why separate considerations are applied in the
present paper to particular $\rho$ and/or $\omega$ decays.

\subsection{Decays $\,(\omega,\rho)\to\pi^+\pi^-$}

The final state $\pi^+\pi^-$ in these decays has isospin $I=1$.
Hence, the standard (and reasonable) assumption is that the direct
amplitude for $\omega^{(0)}\to\pi^+\pi^-$ vanishes (or is very
small), and the decay goes only, or at least mainly, through
mixing\footnote{There are, however, theoretical estimates with
not very small direct $\omega\pi\pi$-transition, see, {\it e.g.}
\cite{WY}.}. Then eq.(15) with $|\wtG_{\omega\rho}|\ll 1$ leads to
\begin{equation}
A(\omega\to\pi^+\pi^-)=
\frac{\wtG_{\omega\rho}}{2}\,A(\rho\to\pi^+\pi^-)\,, ~~~~
\Gamma(\omega\to\pi^+\pi^-)=\frac{|\wtG_{\omega\rho}|^2}{4}\,
\Gamma_\rho\,.
\end{equation}
The present experimental data~\cite{PDT2} lead to
$$\Gamma(\omega\to\pi^+\pi^-) =(144\pm24)~{\rm keV}\,,
~~~\Gamma_\rho=(149.2\pm0.7)~{\rm MeV}$$
and provide
\begin{equation}
|\wtG_{\omega\rho}|=(6.2\pm0.5)\cdot10^{-2}\,,
\end{equation}
in good agreement with qualitative expectations (see the brief
discussion after eq.$(17)\,$). Evidently, the phase of
$\wtG_{\omega\rho}$ can not be determined by using only this
pair of decays.

Together with data \cite{PDT2} on masses and total widths  for $\rho$-
and $\omega$-mesons we obtain
\begin{equation}
|G_{\omega\rho}|=|\wtG_{\omega\rho}\,\delta M^2|=
(3.44\pm0.29)\,10^{-3}~{\rm GeV}^2\,,
\end{equation}
in agreement with phenomenological estimates of other authors
and even with some theoretical estimates.

The small values (21) for $|\wtG_{\omega\rho}|$ and (22)
 for $|G_{\omega\rho}|$ can not, by themselves,
discriminate between different mechanisms of isospin violation
(say, electromagnetic, or related to a definite hadronic channel,
or any other). Phases of those parameters, being extracted from
experimental data, would be very helpful.

One more note is reasonable here. The error for $|\wtG_{\omega\rho}|$
in eq.(21) looks rather small ($<10\%$). However, the true uncertainty
seems to be higher. {\it E.g.}, parameters given in the previous
Particle Data Table~\cite{PDT0} lead to
$$|\wtG_{\omega\rho}|=(7.0\pm0.5)\cdot10^{-2}\,,$$
with deviation about $2\sigma$ from the value (21). That is why
we will use $2\sigma$ level as the uncertainty of
$|\wtG_{\omega\rho}|$ in various numerical estimates below.

\subsection{Decays $\,(\omega,\rho)\to\pi\gamma$}

Isospin conservation does not forbid the direct transitions,
both $\rho^{(0)}\to\pi^0\gamma$ and $\omega^{(0)}\to\pi^0\gamma$,
since $\gamma$-quantum has two isospin components. Therefore, we
need some information on relation between the two amplitudes. The
corresponding exact predictions are still absent, and some models
should be used.  Here we will apply relations \begin{equation}
A^{(0)}(\omega\to\pi^0\gamma)=3\,A^{(0)}(\rho^0\to\pi^0\gamma)
=3\,A^{(0)}(\rho^\pm\to\pi^\pm\gamma)\,,
\end{equation}
that were predicted years ago \cite{Az,An,Sol,BM} on the basis of the
quark model (in the form known at present as the additive quark model).
They were derived with two main assumptions: 1)~mesons consist of one
quark-antiquark pair; 2)~quark charges have their conventional values.
In particular, the coefficient 3 is really a combination of charges:
$$3=\frac{e_u-e_d}{e_u+e_d}\,.$$
Note also that eq.(23) needs a special choice of the relative phase
between $\omega^{(0)}$ and $\rho^{(0)}$. As a matter of fact, the
phase was fixed by standard assumptions that $\omega^{(0)}=(u\ovl
u+d\ovl d)/\sqrt{2},~~\rho^{(0)} =(u\ovl u-d\ovl d)/\sqrt{2}$.

More refined approaches, like QCD sum rules, lead to nearly the same
relations, but with much more complicated derivations, which become
sometimes a kind of art. The limit $N_c\to\infty$, suggested in
ref.\cite{WJY} as a basis for relation (23), is not adequate. It does
provide mesons containing only one quark-antiquark pair, but quark
charges should be \mbox{$N_c$-dependent} to prevent the triangle
anomaly in the Standard Model. Hence, this limit would give
different coefficients for eq.(23). (More detailed discussion
of difficulties of the Standard Model at $N_c\to\infty$ see in
ref.\cite{Sh}.)

The $(\rho\omega)$-mixing does not affect the decay
$\rho^\pm\to\pi^\pm\gamma$, and we can compare it with other decays
to check predictions of the mixing picture.

If relations (23) are correct, the physical amplitude
for $\omega\to\pi^0\gamma$ is practically the same as
$A^{(0)}(\omega\to\pi^0\gamma)$, and the ratio of
widths for  $\omega\to\pi^0\gamma$ and
$\rho^\pm\to\pi^\pm\gamma$
\begin{equation}
r_{\omega/\rho\pm\pi}\equiv
\frac{\Gamma(\omega\to\pi^0\gamma)}
{\Gamma(\rho^\pm\to\pi^\pm\gamma)} =
9\left|1+\frac16\,\wtG_{\omega\rho}\right|^2
\end{equation}
is nearly independent of the mixing.

Present experimental data \cite{PDT2} give
$$\Gamma(\omega\to\pi^0\gamma)=(734\pm34)~{\rm keV},~~~
\Gamma(\rho^\pm\to\pi^\pm\gamma)=(67.1\pm7.5)~{\rm keV},$$
\begin{equation}
r_{\omega/\rho\pm\pi}
=(10.9 \pm 1.3)\,.
\end{equation}
This value reasonably agrees with the "bare" expectation of 9.
If the deviation from 9 is, nevertheless, definitely confirmed,
it could be a result of $(\omega\rho)$-mixing. However, such
possibility looks doubtful, since the mixing correction in eq.(24)
with the value (21) can not exceed 3\%. Furthermore,
the mixing interpretation of the value (25) requires
Re$\,\wtG_{\omega\rho}>0 \,$, while other decay data, more sensitive
to mixing, prefer Re$\,\wtG_{\omega\rho}<0$ (see below). Therefore,
more reasonable would be to admit deviation of the coefficient in
eq.(23) from 3. Taking literally, the value (25) without mixing
leads to 3.3 instead of 3. Note that increase of this coefficient
would diminish the coefficient before $\wtG_{\omega\rho}$ in
eq.(24), 1/6.6 instead of 1/6, and, hence, would diminish the mixing
influence on the decay $\omega\to\pi^0\gamma\,$.

Neutral decay mode $\rho^0\to\pi^0\gamma$ should be stronger
affected by mixing. Combining eqs.(15),(23), we obtain its relative
width in respect to $\rho^{\pm}\to\pi^{\pm}\gamma$ in the form
\begin{equation}
r_{\rho0/\rho\pm\pi}\equiv\frac{\Gamma(\rho^0\to\pi^0\gamma)}
{\Gamma(\rho^\pm\to\pi^\pm\gamma)} =
\left|1-\frac32\,\wtG_{\rho\omega}\right|^2\,.
\end{equation}
Now we can apply $T$-invariance and use the value (21).
Nevertheless, because of unknown phase of the mixing parameter
one can determine only boundaries for $r_{\rho0/\rho\pm\pi}$,
but not its value. With possible $2\sigma$ deviation
for $|\wtG_{\rho\omega}|$  we obtain
\begin{equation}
0.80\leq r_{\rho0/\rho\pm\pi}\leq1.23\,.
\end{equation}
Then the current value Br$\,(\rho^{\pm}\to\pi^{\pm}\gamma)=
(4.5\pm0.5)\cdot10^{-4}$ \cite{PDT2} gives
\begin{equation}
3.2\cdot10^{-4}\leq{\rm Br}\,
(\rho^0\to\pi^0\gamma)\leq6.1\cdot10^{-4}\,.
\end{equation}
Now, if we knew the reliable experimental value of
$r_{\rho0/\rho\pm\pi}$, we would be able to separate
Re$\,\wtG_{\rho\omega}$ and $|{\rm Im}\,\wtG_{\rho\omega}|$
in addition to the value (21) for $|\wtG_{\rho\omega}|$. Note
that higher/lower values of $r_{\rho0/\rho\pm\pi}$ correspond
to negative/positive values of Re$\,\wtG_{\rho\omega}\,$.
Thus, enhancement/suppression of $\rho^0\to\pi^0\gamma$
in respect to $\rho^{\pm}\to\pi^{\pm}\gamma$ implies
negative/positive sign of Re$\,\wtG_{\rho\omega}\,$.

It is worth to emphasize that such correlation is totally independent
of the exact value of the coefficient in eq.(23). Therefore, even
not too accurate experimental comparison of neutral and charged
modes of $\rho\to\pi\gamma\,$ directly and reliably measures
the sign of Re$\,\wtG_{\rho\omega}\,$.

If one neglects the mixing influence on Br$\,(\omega\to
\pi^0\gamma)$ and uses the empirical value (25) to correct
the coefficient in eq.(23), then the coefficient before
$\wtG_{\rho\omega}$ in eq.(26) increases, 3.3/2 instead of 3/2,
thus increasing the mixing influence on $\rho^0\to\pi^0\gamma\,$.
Numerically, however, boundaries of intervals (27),(28) stay
nearly the same.

We can also construct one more ratio
\begin{equation}
r_{\omega/\rho0\pi}\equiv\frac{\Gamma(\omega\to\pi^0\gamma)}
{\Gamma(\rho^0\to\pi^0\gamma)} =
9\,\left|\frac{1+\frac16\,\wtG_{\omega\rho}}
{1-\frac32\,\wtG_{\rho\omega}}\right|^2
\end{equation}
with boundaries
\begin{equation}
7.2\leq r_{\omega/\rho0\pi}\leq11.6\,.
\end{equation}
Note that lower/upper boundary in eq.(30) corresponds to
upper/lower boundaries for intervals (27),(28) and to
negative/positive sign of Re$\,\wtG_{\rho\omega}$.

Theoretical estimations for $\rho^0\to\pi^0\gamma\,$, as a rule,
agree with the phenomenological intervals (27),(28), with
tendency to their upper ends (see, {\it e.g.}, ref.\cite{BOC}).
The only exclusion is the essentially higher estimate~\cite{WJY}.
It is interesting to trace the source of such deviation. Detailed
comparison shows that instead of
$$-\frac12\,\wtG_{\rho\omega}=\frac{G_{\rho\omega}}
{M^2_\rho-M^2_\omega} $$
the amplitude of ref.\cite{WJY} contains the quantity
$$-\frac12\,\wtG'_{\rho\omega}=\frac{G_{\rho\omega}}
{m^2_\rho-m^2_\omega+i\,m_\omega\Gamma_\omega }$$
with the same value of $G_{\rho\omega}$ (up to uncertainties and
notations). At the current values of masses and widths~\cite{PDT2}
this provides the additional enhancing factor
$$
\left|\frac{M^2_\rho-M^2_\omega}
{m^2_\rho-m^2_\omega+i\,m_\omega\Gamma_\omega}
\right|=
\left|\frac{m^2_\rho-m^2_\omega-i\,m_\rho\Gamma_\rho
+i\,m_\omega\Gamma_\omega}{m^2_\rho-m^2_\omega+
i\,m_\omega\Gamma_\omega }\right|=5.8\,,$$
which transforms the intervals (27),(28) into
$$
0.14\leq r_{\rho0/\rho\pm\pi}\leq2.65\,,~~~~
0.56\cdot10^{-4}\leq{\rm Br}\,
(\rho^0\to\pi^0\gamma)\leq13.25\cdot10^{-4}\,.$$
The upper ends here just agree with the estimates of ref.\cite{WJY}.
However, derivation in the previous section demonstrates that
the mixing parameter for production and decay amplitudes
(see expressions $(14),(15)\,$) should contain in its denominator
the difference of pole masses, even though $k^2$ in propagators
takes only real values and does not reach any of the pole (complex)
masses.

Let us discuss the experimental situation. The latest version of
Particle Data Tables~\cite{PDT2} gives the value
$${\rm Br}\,(\rho^0\to\pi^0\gamma)=(7.9\pm2.0)\cdot10^{-4}\,,$$
based on one experiment~\cite{Dol} only. Reanalysis of all
existing data on $e^+e^-\to\pi^0\gamma$ was presented in~\cite{BEI}
with taking into account coherent contributions of various
resonances. It provided, with some model assumptions, two sets of
acceptable solutions for Br$\,(\rho^0\to\pi^0\gamma)$, one between
$6\cdot10^{-4}$ and $7\cdot10^{-4}$, another between $11\cdot10^{-4}$
and $12\cdot10^{-4}$, all higher than Br$\,(\rho^{\pm}\to\pi^{\pm}
\gamma)=(4.5\pm0.5)\cdot10^{-4}$~\cite{PDT2}. The previous version
of Particle Data Tables~\cite{PDT0} used the lower solution for a
particular model, though ref.\cite{BEI} gave only meager motivations
for this model and this solution. There are arguments showing that
the results~\cite{BEI} for $\rho^0\to\pi^0\gamma$ are still rather
uncertain: the models used assume non-PDG values of $m_\rho$ and/or
$\Gamma_\rho\,$; the triangle anomaly contribution is assumed to be
the only non-resonant background, but the out-of-resonance data can
not confirm its presence in $e^+e^-\to\pi^0\gamma$ (though do confirm
the similar anomaly contribution to $e^+e^-\to\eta\gamma$); phase
relations between various resonance contributions look strange and
unexpected. The own conclusion of the authors of ref.~\cite{BEI} is
that more measurements, with better accuracy, are necessary for
the $\pi^0\gamma$ final state to obtain a firm result.

Meanwhile, the above value used in the latest Tables~\cite{PDT2} looks
acceptable at the moment, just due to its large error. Though with
such or even larger uncertainties, all published measurements give
evidence for the enhancement of $\rho^0\to\pi^0\gamma$ in respect to
$\rho^{\pm}\to\pi^{\pm}\gamma$ and, thus, evidence for the negative
sign of Re$\,\wtG_{\rho\omega}$.

\subsection{Decays $\,(\omega,\rho)\to\eta\gamma$}

Assumptions, which lead to relations (23), provide similar
relations also for amplitudes of some other decays. {\it E.g.}, for
decays to $\eta\gamma$ we obtain
\begin{equation}
3\,A^{(0)}(\omega\to\eta\gamma)=A^{(0)}(\rho^0\to\eta\gamma)\,.
\end{equation}
The factor 3 has here the same nature as in eq.(23), though it
makes more intensive (surely, in terms of partial widths, not
of branchings) decay of $\rho^0$ instead of $\omega$.

For the final state $\eta\gamma$ we have no analog of the modes
$\rho^{\pm}\to\pi^{\pm}\gamma$, insensitive to the
$(\rho\omega)$-mixing. Nevertheless, in analogy with the ratio
$r_{\omega/\rho0\pi}$ of eq.(29), we can construct another ratio
\begin{equation}
r_{\rho0/\omega\eta}\equiv\frac{\Gamma(\rho^0\to\eta\gamma)}
{\Gamma(\omega\to\eta\gamma)}=9\,\left|\frac{1-\frac16\,
\wtG_{\rho\omega}}{1+\frac32\,\wtG_{\omega\rho}}\right|^2\,.
\end{equation}
With $2\sigma$ boundaries for $|\wtG|$ it has admissible
interval
\begin{equation} 7.2\leq r_{\rho0/\omega\eta}\leq11.6\,,
\end{equation}
numerically the same as in eq.(30). Note, however, different
correlations: the lower/upper boundary in eq.(33) corresponds
to the lower/upper boundaries in eqs.(27),(28), but to the
upper/lower boundary in eq.(30). In terms of Re$\,\wtG$
the lower/upper boundary in eq.(33) corresponds to
positive/negative Re$\,\wtG\,$, opposite to eq.(30).

Let us consider the present experimental situation. Particle Data
Group~\cite{PDT2} gives, after all evaluations,
\begin{equation}
{\rm Br}\,(\rho^0\to\eta\gamma) =(3.8 \pm 0.7) \cdot 10^{-4}\,,~~~
\Gamma(\rho^0\to\eta\gamma) =(57 \pm 10)~{\rm keV}
\end{equation}
and
\begin{equation}
{\rm Br}\,(\omega\to\eta\gamma) =(6.5 \pm 1.1)\cdot 10^{-4}\,,~~~
\Gamma(\omega\to\eta\gamma) =(5.5 \pm 0.9)~{\rm keV}\,.
\end{equation}
This implies
\begin{equation}
r_{\rho0/\omega\eta}=10.3 \pm 2.6\,,
\end{equation}
in agreement with the interval (33). This value
may be considered as an additional evidence for
Re$\,\wtG_{\omega\rho}<0$ and, therefore, as an indirect
evidence for enhancement of $\rho^0\to\pi^0\gamma\,$
due to $(\rho\omega)$-mixing. However, the large error of
the value (36) makes this result rather uncertain.

\subsection{Decays $\,\eta'\to(\omega,\rho)\,\gamma$}

Bare amplitudes of these decays are related just as in decays with
$\eta$-meson:
\begin{equation}
3\,A^{(0)}(\eta'\to\omega\gamma)=A^{(0)}(\eta'\to\rho^0\gamma)\,.
\end{equation}
Therefore, similar to $r_{\rho0/\omega\eta}\,$,
we can construct the ratio
\begin{equation}
r_{\eta'\rho0/\omega}\equiv\frac{\Gamma(\eta'\to\rho^0\gamma)}
{\Gamma(\eta'\to\omega\gamma)}=
9\,\left|\frac{1-\frac16\,\wtG_{\omega\rho}}
{1+\frac32\,\wtG_{\rho\omega}}\right|^2
\end{equation}
with the same boundaries
\begin{equation}
7.2\leq r_{\eta'\rho0/\omega}\leq11.6\,.
\end{equation}
Its correlations with other similar ratios are also the same
as for $r_{\rho0/\omega\eta}\,$.

Experimental data \cite{PDT2} give
\begin{equation}
{\rm Br}\,(\eta'\to\rho^0\gamma) =(29.5 \pm 1.0)\%\,,~~~
{\rm Br}\,(\eta'\to\omega\gamma) =(3.03 \pm 0.31)\%\,,
\end{equation}
that lead to the value
\begin{equation}
r_{\eta'\rho0/\omega}=(9.74 \pm 1.05)
\end{equation}
inside the expected interval (39). It looks to be shifted upward
from 9, thus giving evidence for Re$\,\wtG_{\rho\omega}<0$ and the
enhanced decay $\rho^0\to\pi^0\gamma$.  But such small shift with
rather large error still prevents one from any firm conclusion.

\subsection{Decays $\,(\omega,\rho)\to e^+e^-$}

Decay of a neutral $C$-odd vector meson to $e^+e^-$ pair goes
through one virtual photon. If the meson may be considered to
consist of a quark-antiquark pair, the decay amplitude should be
equal to the quark charge $e_q$ multiplied by some hadronic matrix
element.  (In terms of the constituent quark picture it is proportional
to the short-distance value of the internal wave function.)

Situation is somewhat different for $\omega$ and $\rho$. Here,
even for bare states, $\omega^{(0)}$ and $\rho^{(0)}\,$, we have
coherent mixtures of at least two flavours with different charges:
$(\ovl uu+\ovl dd)/\sqrt{2}$ and \mbox{$(\ovl uu-\ovl dd)/\sqrt{2}\,$}.
Here we can use some averaged charges as the effective charges
$e_\omega$ and $e_\rho\,$.

If direct isospin violation is absent (or may be neglected), so that
the arising hadronic matrix elements are the same for $\ovl uu$
and $\ovl dd$ components, then annihilation of the bare states
(transforming them into vacuum) by the quark e.m. current provides
the effective charges in the form
$$e_\omega=\frac{e_u+e_d}{\sqrt{2}}= \frac1{3\sqrt{2}}\,,~~~~
e_\rho=\frac{e_u-e_d}{\sqrt{2}}=\frac1{\sqrt{2}}\,.$$
Note the relative factor 3 that appears here again. It is natural,
therefore, to expect that bare decay amplitudes satisfy relations
similar to eqs.(31),(37):
\begin{equation}
3\,A^{(0)}(\omega^{(0)}\to e^+e^-)=A^{(0)}(\rho^{(0)}\to e^+e^-)\,.
\end{equation}
Thus, in full similarity to previous cases, one can construct the
ratio for physical quantities
\begin{equation}
r_{\rho0/\omega(ee)}\equiv\frac{\Gamma(\rho^{(0)}\to e^+e^-)}
{\Gamma(\omega^{(0)}\to e^+e^-)}=
9\,\left|\frac{1-\frac16\,\wtG_{\rho\omega}}
{1+\frac32\,\wtG_{\omega\rho}}\right|^2\,,
\end{equation}
again with the same boundaries
\begin{equation}
7.2\leq r_{\rho0/\omega(ee)}\leq11.6\,.
\end{equation}
and the same correlations with other similar ratios and with
the sign of Re$\,\wtG$ as for $r_{\rho0/\omega\eta}\,$ or
$r_{\eta'\rho0/\omega}\,$. According to present
experimental data~\cite{PDT2}
$$\Gamma(\rho^0\to e+e-)=(6.85 \pm 0.11) {\rm keV}\,,~~~~
\Gamma(\omega\to e+e-)=(0.60 \pm 0.02) {\rm keV}\,,$$
\begin{equation}
r_{\rho0/\omega(ee)}=(11.42 \pm 0.42)\,.
\end{equation}
This value reasonably agrees with the values (36),(41) and gives
a clearer evidence for Re$\,\wtG_{\omega\rho}<0\,$.

Decays to $e^+e^-$ seem to admit even more detailed test
of the mixing picture. For heavy quarkonia, where hypothesis of
one quark-antiquark pair looks fulfilled, there is an empirical
observation that the partial widths of their decays to $e^+e^-$
equals just a constant multiplied by the corresponding quark
charge squared:
\begin{equation}
\Gamma(Q\ovl Q\to e^+e^-)=e_Q^2\,\Gamma_0\,.
\end{equation}
Indeed, let us consider three heavier quarkonia,  $\Upsilon,\,
J/\psi,\,\varphi$  corresponding (with good accuracy) to the
definite flavour of the constituent quark (and antiquark) and,
hence, to the definite value of $e_Q\,$:  \mbox{$e_b=-1/3,\,$}
$e_c=2/3,\,e_s=-1/3\,$. Then, from experimental data~\cite{PDT2},
\begin{equation}
\Gamma_0^{(\Upsilon)}=(11.88\pm0.45)~{\rm keV}\,,
~~~ \Gamma_0^{(J/\psi)}=(11.84\pm0.83)~{\rm keV}\,,
~~~ \Gamma_0^{(\varphi)}=(11.34\pm0.18)~{\rm keV}\,.
\end{equation}
We can try to check this regularity for $\rho^0,\,\omega$
as well, using effective charges $e_\rho,\,e_\omega\,$. Then
the data~\cite{PDT2} lead to values
\begin{equation}
e_\omega^{-2}\,\Gamma(\omega\to e^+e^-)=10.80\pm0.36~{\rm keV}\,,
~~~~e_\rho^{-2}\,\Gamma(\rho^0\to e^+e^-)=13.70\pm0.22~{\rm keV}\,,
\end{equation}
which look somewhat lower (for $\omega$) or higher (for $\rho$)
than the "normal" values (47).

There are at least three
possible explanations: 1) insufficient precision prevents from
any statement of differences between numerical values (47) and
(48); 2) the present level of understanding QCD is not sufficient
for extrapolating the properties of heavy quarkonia to lighter ones;
3) values (48) for physical mesons $\rho,\,\omega$ may deviate from
(47) due to mixing of bare states $\rho^{(0)},\,\omega^{(0)}\,$.
The first two points imply that any serious discussion should be
postponed till further experimental and/or theoretical progress.
Therefore, we will not touch them here and now; instead we restrict
ourselves to the third possibility.

Let us assume that the above regularity would be correct
for the bare states $\rho^{(0)},\,\omega^{(0)}$  and
that $\Gamma_0$ is indeed a universal quantity (without
discussing why). Then the $(\rho\omega)$-mixing changes
widths for the physical states $\rho^0,\,\omega\,$ so that
\begin{equation} e_\omega^{-2}\,\Gamma(\omega\to
e^+e^-)=\Gamma_0 \left|1+\frac32\,\wtG_{\omega\rho}\right|^2\,,
~~~e_\rho^{-2}\,\Gamma(\rho^0\to e^+e^-)=\Gamma_0
\left|1-\frac16\,\wtG_{\rho\omega}\right|^2\,.
\end{equation}
In such a framework the relation of numerical values (48),
for $\rho^0$ higher than for $\omega$, gives a new strong
support to Re$\,\wtG_{\rho\omega}<0\,$ and, hence, to
enhancement of the mode $\rho^0\to\pi^0\gamma$ in respect
to $\rho^{\pm}\to\pi^{\pm}\gamma\,$.

Further, taking for definiteness the heavier quarkonium
value $\Gamma_0=11.86~{\rm keV}\,$ from eq.(47), we obtain
expected intervals
\begin{equation} 9.44~{\rm keV}\leq e_\omega^{-2}\,
\Gamma(\omega\to e^+e^-)\leq14.56~{\rm keV} \end{equation} and
\begin{equation} 11.58~{\rm keV}\leq e_\rho^{-2}\, \Gamma(\rho^0
\to e^+e^-)\leq12.15~{\rm keV} \end{equation}
Quantitatively, the value (48) for $\omega$ is in good agreement
with the interval (50), while the value for $\rho^0$ noticeably
exceeds the expected upper boundary (51). This could mean either
that $\Gamma_0^{(\rho,\,\omega)}$ deviates from
$\Gamma_0^{(\Upsilon)}\approx\Gamma_0^{(J/\psi)}$ or even that
$\Gamma_0^{(\rho)}\ne\Gamma_0^{(\omega)}$ due, {\it e.g.}, to direct
isospin violation for decay amplitudes of the bare states (see
discussion below). Having in mind the universal character of the
mixing parameter $\wtG_{\rho\omega}=\wtG_{\omega\rho}\,$, one
may hope that precise investigation of a wider set of processes
will allow to clarify the situation.

One may add here one more notice. Of course, all three values
(47) coincide at the level not worse than $1\sigma\,$. Nevertheless,
$\Gamma_0^{(\Upsilon)}$ and $\Gamma_0^{(J/\psi)}$ are equal to each
other with much better accuracy, while $\Gamma_0^{(\varphi)}$ is
somewhat lower. Such situation, if confirmed, could be due to mixing
of $\varphi$ with $\omega$ and/or other mesons.

\section {Discussion}

As was demonstrated in the preceding section, the
$(\rho\omega)$-mixing manifests itself not only in the
well-known decay $\omega\to\pi^+\pi^-$ and in radiative decay
$\rho^0\to\pi^0\gamma$, but also in some other electromagnetic
decays with participation of $\rho$ or $\omega$, in either initial
or final state. Particular modes of interest are radiative decays
$(\rho^0,\omega)\to\eta\gamma,~\eta'\to(\rho^0,\omega)\gamma$ and
decays $(\rho^0,\omega)\to e^+e^-$ going through one virtual photon.
The central point of studies appears to be a special consistent
correlation between properties of decays in various pairs.

Existing data for the decay pairs may be presented on the complex plane
of the mixing parameter $\wtG$, as seen at fig.1. If the role of mixing
for the decays is correctly described in the preceding section, then
all the corresponding bands should overlap. The presently achieved
accuracy is not yet sufficiently informative. However, the data do not
contradict to overlapping at Re$\,\wtG<0$, which corresponds to
enhancement of $\rho^0\to\pi^0\gamma$ in respect to
$\rho^{\pm}\to\pi^{\pm}\gamma\,$. Being done with
better precision, experiments on those (and other) decays could
check the expected correlation of properties of different processes
and, thus, confirm (or reject) the role of mixing.

Let us analyze the nature of that correlation. The first step
in its description begins with relations for bare (unmixed)
amplitudes. At first sight, the used relations
(23),(31),(37),(42) may be justified only in the framework of a
specific (constituent quark) model. However, they have a more
general origin. Indeed, electromagnetic interactions of hadrons
in the quark picture are manifestations of the "microscopic"
interaction proportional (for light quarks) to
$$e_u\,\ovl uu\,+e_d\,\ovl dd\,=\frac{e_u+e_d}{\sqrt2}\,
\frac{\ovl uu+\ovl dd}{\sqrt2}+\frac{e_u-e_d}{\sqrt2}\,
\frac{\ovl uu-\ovl dd}{\sqrt2}$$
(of course, we mean the vector current structure,
well-known and not shown here explicitly in detail). Hence,
the canonical quark charges imply that isovector component
of the photon is coupled to light quarks 3 times stronger
than isoscalar one.

All the considered pairs of decays, differing by interchange
$\rho^0\rlh\omega$, have a common property: one of them
contains only isovector component of the photon (real or
virtual), while only isoscalar component of the photon
participates in another decay. Relations (23),(31),(37),(42)
correspond to the simple expectation from the above discussion
that amplitudes for light-quark processes with the isovector
photon are 3 times larger than that for similar processes with
isoscalar photon. Of course, these simple relations may be modified
in particular processes by specific hadronic matrix elements.
Nevertheless, one may argue that the modifications should not be
large.

Indeed, the processes discussed here are soft, and essential
contributions to their amplitudes come from the photon
coupling to valence quarks. Now, since the valence quark
structure inside any hadron is similar to the constituent
quark one, it is natural to expect that relations for bare
amplitudes are closely similar to the used relations
(23),(31),(37),(42). Such arguments seem to be applicable
both for radiative decays and for $e^+e^-$-decays of mesons.
Note that similar reasoning might also explain why QCD
calculations (say, the sum rules) and constituent quarks
provide nearly the same predictions for meson radiative decays.

The above relations between processes with isovector
{\it vs.} isoscalar photon might be applicable to amplitudes
for "bare" (unmixed) states, where isospin could be a good
quantum number. Then the next step  should be description of
isospin violation by the $(\rho\omega)$-mixing. It makes the
relations for physical (mixed) amplitudes of the decays
to be somewhat modified in comparison with ones for bare
amplitudes. Essential point is that different decay pairs
are modified in a correlated way, since in all cases the
mixing is described by the same universal dimensionless
(generally, complex) phenomenological parameter
\mbox{$\wtG_{\rho\omega}\,(=\wtG_{\omega\rho})\,$.}

Future accurate experiments should allow to check whether
all those correlations are correct and, thus, examine
consistency of the picture. But some piece of information
does exist even now. Data on $\omega\to\pi^+\pi^-$ allow
to find the absolute value of the mixing parameter
$|\wtG_{\rho\omega}|$. If the decays
$(\rho^0,\omega)\to\pi^0\gamma$ were measured at least
with the same precision as $\rho^{\pm}\to\pi^{\pm}\gamma$,
we could extract also Re$\,\wtG_{\rho\omega}$
and then test hypotheses on the mechanism of the
$(\rho\omega)$-mixing.

Meanwhile, existing direct measurements give evidence for
enhancement of $\rho^0\to\pi^0\gamma$ in comparison with
$\rho^{\pm}\to\pi^{\pm}\gamma$ (exact number is still
to be determined). This implies that
Re$\,\wtG_{\rho\omega}<0$ and suggests a special kind for
modification of amplitudes in other pairs of decays
with participation of $\rho^0,\omega$.

As was demonstrated in the preceding Section and in fig.1, current data
on $(\rho^0,\omega)\to\eta\gamma,\, \eta'\to(\rho^0,\omega)\gamma$ and
$(\rho^0,\omega) \to e^+e^-$ are not confirmative yet, but nevertheless
give additional support for negative Re$\,\wtG_{\rho\omega}$ and,
therefore, indirectly confirm enhancement of $\rho^0\to\pi^0\gamma\,$.
Since $\,\wtG_{\rho\omega}=G_{\rho\omega}/\delta M^2$ with nearly
imaginary $\delta M^2=(M_\omega^2-M_\rho^2)/2$, this implies
also that the direct $(\rho\omega)$-vertex $G_{\rho\omega}$
is not real and, as a result, may reject even now some
simplified models for the $(\rho\omega)$-transition.

General character of the used relation between isovector and
isoscalar components of the photon may be checked by testing
it in a wider set of decays after they become accessible. As
an example we can take the pair of decays
$\,(\rho^0,\omega)\to\pi^0\pi^0\gamma\,$, where the photon
has $I=1$ for $\rho^0$-decay and $I=0$ for $\omega$-decay.
Particle Data Tables~\cite{PDT2} give
\begin{equation} {\rm Br}\,(\rho^0\to\pi^0\pi^0\gamma)=
(4.8^{+3.4}_{-1.9})\cdot 10^{-5}\,,~~~~{\rm Br}\,
(\omega\to\pi^0\pi^0\gamma)=(7.8\pm3.4)\cdot10^{-5}\,.
\end{equation}
Together with data on total widths we obtain
\begin{equation}
r_{\rho0/\omega(\pi0\pi0)}\equiv \frac{\Gamma(\rho^0\to
\pi^0\pi^0\gamma)}{\Gamma(\omega\to\pi^0\pi^0\gamma)}
\approx11\,.
\end{equation}
Large experimental uncertainty of branchings (52) prevents
us from more detailed discussion of these decays. However,
they may be useful and helpful in future studies. But even
at present one can notice close equality of
$r_{\rho0/\omega(\pi0\pi0)}$ to other $r$-ratios of the
previous Section. This confirms universality of stronger
isovector {\it vs.} isoscalar interaction for the photon,
just at the expected quantitative level.

All numerical estimations in this paper have been
made under assumption that all necessary parameters
are constant. Those parameters are $\rho,\,\omega$ complex
masses ({\it i.e.}, masses and widths) and mixing parameters
$G$ (or $\wtG$). Such approach is analogous to the standard
Breit-Wigner description of a resonance amplitude, with fixed
values of energy (mass) and width. It is known to work quite
well for description of narrow peaks, as $\omega$ or $\varphi$.
However, to describe the broad $\rho$-peak one needs to account
for $k^2$-dependence of, at least, the $\rho$-width. Moreover,
even to describe the $(\varphi\omega)$-interference in $e^+e^-
\to\pi^+\pi^-\pi^0$ one seems to need the "long tail" of the
$\omega$-resonance, with accounting for $k^2$-dependence of its
width~\cite{AK}. These examples show that, most probably, detailed
description of, say, $e^+e^-\to\pi^0\gamma$ for extraction of the
partial $\rho$-width and $(\rho\omega)$-interference may require to
consider the $k^2$-dependence of parameters (at least, at
future level of precision).

Another simplifying hypothesis used was the leading role of the
$(\rho\omega)$-mixing for isospin violation. A simple structure
was assumed for bare amplitudes, corresponding to "minimal"
violation of isospin\footnote{Since all the discussed decays,
except may be $(\rho,\omega)\to\pi^+\pi^-\,$, are evidently
electromagnetic, they surely violate isospin. But the violation
accounted for was only due to difference of charges $e_u,\,e_d$,
without account for difference of quark masses or other properties.}.
There are, however, arguments for necessity of more complicated
structure, with direct violation of the isospin in
bare amplitudes.

Indeed, let us consider first the radiative decays. In terms of
the constituent quarks their transition amplitudes are determined
by magnetic moments of the quarks, which manifest themselves also
in baryon magnetic moments. The factor 3 in relations
(23),(31),(37) corresponds to the assumption that magnetic
moments of $u,\,d$ are equal to their charges $e_u,\,e_d$
multiplied by the same factor. Since this assumption implies also
the ratio of the proton/neutron magnetic moments $\mu_p/\mu_n=-3/2$,
we know that it is only approximate. The well established (small)
deviation of this magnetic moments ratio from -3/2 gives evidence
for difference of factors in the quark magnetic moments (the same
conclusion comes from magnetic moments of other baryons), due to
different masses or because of other reasons\footnote{It is
interesting to note that the corresponding factor for the heavier
$d$-quark appears 5\% larger than for lighter $u$-quark, contrary
to familiar properties of normal magnetic moments. This can be
viewed as evidence that (constituent) quarks may have anomalous
magnetic moments.}. Thus, the factor 3 should be corrected, and
the corrections can be extracted from existing data\footnote{In
framework of the constituent quark model it is 3.21 instead of 3
in eqs.(23),(31),(37), which gives the factor 10.3 instead of 9
for $r$'s, in agreement with the present experimental value (25).}.
However, repeating the analysis of Section 3 with these corrections
shows that today they appear to be corrections to corrections
in comparison with effects of $(\rho\omega)$-mixing.

Arguments for direct isospin violation in $e^+e^-$-decays look
differently. It is essential here that both $\rho^{(0)}$ and
$\omega^{(0)}$ have two flavour components, their couplings to
photon being proportional just to charges $e_u,\,e_d\,$. Thus,
each of the bare decay amplitudes is a combination of two flavour
contributions.  According to the constituent quark model, every
contribution due to annihilation of a pair $\ovl QQ$ is proportional
to the product of $e_Q$ and the corresponding short-range wave
function $\psi_{Q}(0)$.  Exact isospin conservation implies equality
$\psi_{u}(0)=\psi_{d}(0)$.  However, the scaling relation (46) leads
to the phenomenological dependence on the quark mass\cite{AFK}
$$|\psi_{Q}(0)|^2\propto m_Q^2$$ (note that it would be $m_Q^3$
for the Coulomb-like potential).  Now, inequality of $m_u$ and
$m_d$ should influence the amplitudes $\rho^{(0)}\to e^+e^-,\,
\omega^{(0)}\to e^+e^-$ and deviate their ratio from 3 (recall
that we deal here with constituent quarks, so the correction
should be at a level of several percents, not several times
as it would be for current quark masses)\footnote{ Interestingly
enough, this mechanism acts in the same direction as mixing:  it
enhances $\rho^{(0)}\to e^+e^-$ and suppresses $\omega^{(0)}\to
e^+e^-\,$, thus increasing the coefficient in eq.(42). Such changes
are favourable, since they shift the theoretically expected intervals
(50),(51) just so to adjust  them to experimental values (48).}.

Of course, there are also some other corrections. {\it E.g.}, $\,\omega$
contains an admixture of strange quarks which may be described
as mixing of $\omega$ and $\varphi$ with mixing angle
$\alpha_V\approx4^{\circ}$. Corresponding relative corrections
for decays of Section 3 are of order $\sim\sin^2
\alpha_V\approx5\cdot10^{-3}$. Their influence appears even smaller
than discussed above and may be necessary only at future levels of
precision.

\section {Conclusions}

Results of the present paper may be briefly summarized as follows.
\begin{enumerate}
\item Independently of a framework of any effective field theory,
the $(\rho\omega)$-mixing is completely determined by two universal
parameters $\wtG_{\rho\omega}$ and $\wtG_{\omega\rho}$, both
being, generally, complex. For the case of $T$-conservation
(or in the framework of effective field theory) they may be made
equal to each other, staying complex outside the effective field
theory. Experimental determination of the mixing parameter(s)
would allow to check models of isospin violation.
\item  It was known for many years that isospin violation,
due to the $(\rho\omega)$-mixing, should be enhanced in the forbidden
decay $\omega\to \pi^+\pi^-\,$; later the same effect was suggested
for the radiative decay $\rho^0\to\pi^0\gamma$ (its branching ratio
may be unequal to that of $\rho^{\pm}\to\pi^{\pm}\gamma$ due to
interference of the direct decay and cascade decay
$\rho^0\to\omega\to\pi^0\gamma\,$). As shown here, the mixing should
also affect all pairs of decays of $\rho^0,\,\omega$ to the same final
state and decays of heavier particles with production of $\rho^0,\,
\omega\,$.
\item  The $(\rho\omega)$-mixing influences various pairs of
$(\rho^0,\,\omega)$-decays in a regular, correlated manner.
Such regularities agree with existing data on radiative decays
and decays to $e^+e^-$, though achieved precision of data is
insufficient for the firm conclusion. Nevertheless, the data
prefer Re$\,\wtG_{\rho\omega}<0\,$. This, on one side, supports
enhancement of $\rho^0\to\pi^0\gamma$ in comparison with
$\rho^{\pm}\to\pi^{\pm}\gamma\,$ and implies, on the other,
that the direct $(\rho\omega)$-coupling $G_{\rho\omega}=
G_{\omega\rho}$ is complex.
\item The universal nature of the mixing parameter will allow,
at higher experimental accuracy, to separate mixing isospin
violation due to $(\rho\omega)$-transitions from direct isospin
violation in amplitudes of "bare" (unmixed) states $\rho^{(0)}$ and
$\omega^{(0)}\,$. Even present data give some evidence for
necessity of such direct violating effects.
\end{enumerate}

Thus, we can expect that in the near future the meson radiative
decays with participation of $\rho$ and/or $\omega$ may indeed
be attractive and useful for studying the $(\rho\omega)$-mixing
and other manifestations of isospin violation.

About forty years ago, in first years of the quark era,
the radiative decays of mesons were suggested (and really
used) as a mean to check that quarks inside baryons and mesons
are the same~\cite{Az,An,Sol,BM,Th}. Now, forty years later,
at a higher level of experimental precision and theoretical
understanding, such decays may again provide new interesting
information. This time the radiative decays might elucidate
mechanisms of isospin violation.

\section* {Acknowledgements}

I thank V.V.Anisovich for stimulating discussions and V.A.Nikonov
for help in preparing the text. The work was supported in part by
the RFBR grant 00-15-96610.

\newpage
\begin{figure}
\centerline{\epsfig{file=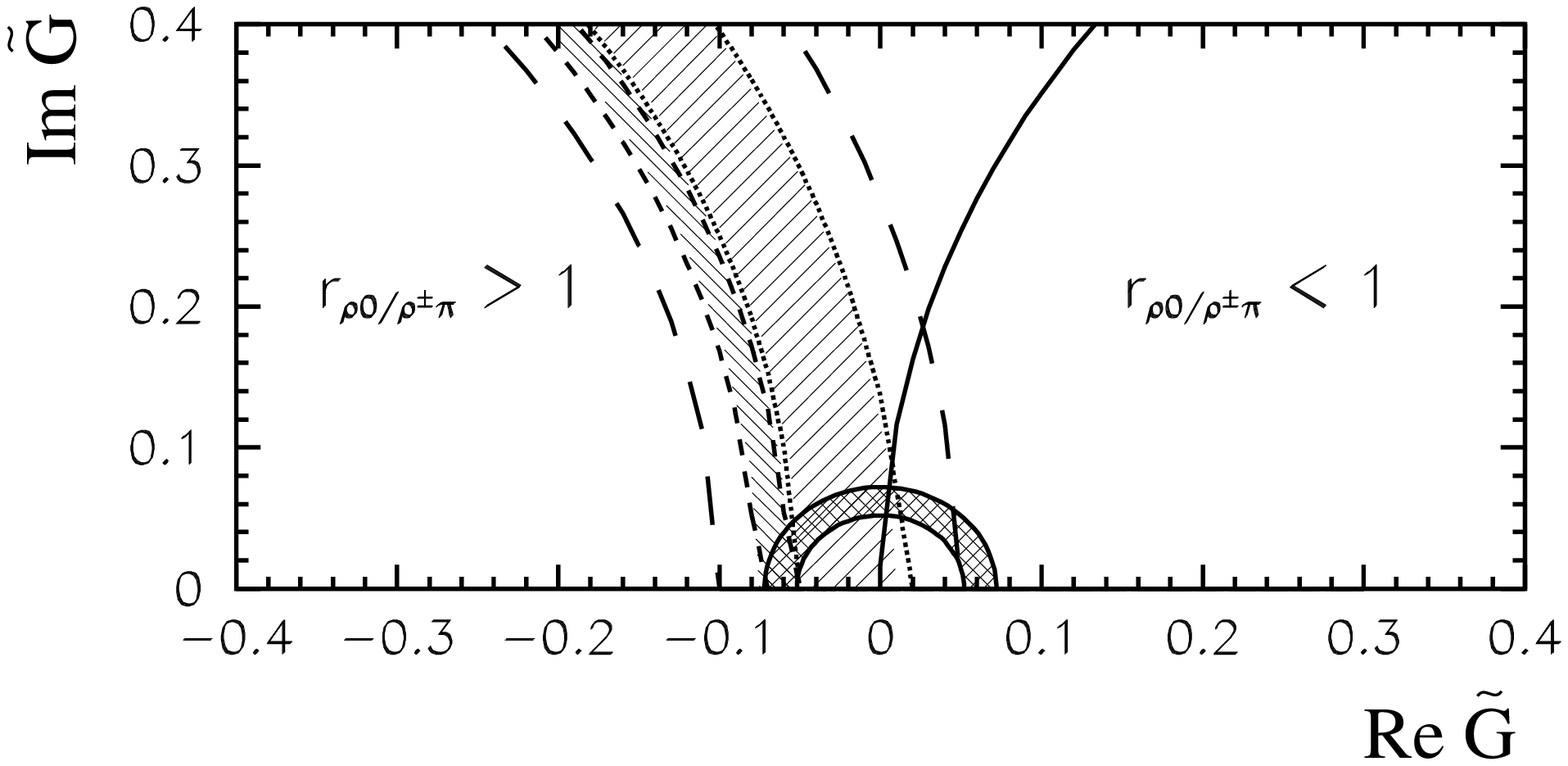,width=16cm}}
\caption{Data on various decay pairs as seen at the complex plane
of $\wtG$. The long-dashed uncovered band is for
$r_{\rho0/\omega\eta}$, eq.(36); the short-dashed
band with left-inclined hatching is for $r_{\rho0/\omega(ee)}$,
eq.(45); the dotted band with right-inclined hatching is for
$r_{\eta'\rho0/\omega}$, eq.(41). The solid ring with double
hatching is for $(\omega,\rho)\to\pi\pi$, eq.(21) with $2\sigma$
width. Space to the left/right of the solid line corresponds
to enhancement/suppression of $\rho^0\to\pi^0\gamma$ in respect
to $\rho^{\pm}\to\pi^{\pm}\gamma$.}
\end{figure}

\end{document}